\documentclass[format=sigconf,anonymous=false,authorversion=false]{acmart}

\usepackage[english]{babel}
\usepackage{booktabs}
\usepackage{listings}
\usepackage{moreverb}
\usepackage{multirow}
\usepackage{graphicx}
\usepackage{subcaption}
\usepackage[flushleft]{threeparttable}
\usepackage[utf8]{inputenc}
\usepackage{xspace}
\usepackage{amssymb}
\usepackage{pifont}  
\usepackage[ruled,vlined,linesnumbered]{algorithm2e}
\usepackage{algpseudocode}

\newcommand{\fig}[3]
{
  \begin{figure}[t]
    \centering
    \includegraphics{fig/#2}
    \caption{#3}
    \label{fig:#2}
  \end{figure}
}

\newcommand{\subfig}[2]
{
  \begin{subfigure}[c]{\linewidth}
    \includegraphics{fig/#1}
    \caption{#2}
    \label{fig:#1}
  \end{subfigure}
}
\newcommand{\fsource}[1]{\footnote{Source: #1}}
\newcommand{\furl}[1]{\fsource{#1}}

\newcommand{\figref}[1]{Figure~\ref{fig:#1}}

\newcommand{\srcref}[1]{Algorithm~\ref{code:#1}}
\newcommand{\tabref}[1]{Table~\ref{tab:#1}}
\newcommand{\secref}[1]{Section~\ref{sec:#1}}

\newcommand{\seetab}[1]{(see \tabref{#1})}
\newcommand{\seesec}[1]{(see \secref{#1})}

\newcommand{\HybridChallengeImprovementVsBaseline}{31.1~\%\xspace}
\newcommand{\HybridChallengeImprovementVsWinner}{12.2~\%\xspace}
\newcommand{\HybridChallengeLeftOutItems}{4140\xspace}
\newcommand{\HybridChallengeNumUsers}{4140\xspace}

\newcommand{\HybridChallengeParamTopThree}{26000\xspace}
\newcommand{\HybridChallengeParamTopTwo}{9000\xspace}

\newcommand{\HybridChallengeRuntimeBestShare}{14.8~\%\xspace}
\newcommand{\HybridChallengeScoreBest}{0.0516\xspace}
\newcommand{\HybridChallengeScoreWinner}{0.0460\xspace}

\newcommand{\HybridMovieLensImprovementVsBaseline}{20.3~\%\xspace}
\newcommand{\HybridMovieLensImprovementVsWinner}{12.8~\%\xspace}
\newcommand{\HybridMovieLensLeftOutItems}{943\xspace}

\newcommand{\HybridMovieLensParamTopThree}{19000\xspace}
\newcommand{\HybridMovieLensParamTopTwo}{5000\xspace}
\newcommand{\HybridMovieLensRepeats}{1000\xspace}
\newcommand{\HybridMovieLensRuntimeBestShare}{19.9~\%\xspace}
\newcommand{\HybridMovieLensScoreBest}{0.0349\xspace}
\newcommand{\HybridMovieLensScoreWinner}{0.0309\xspace}

\copyrightyear{2017}
\acmYear{2017}
\setcopyright{acmlicensed}
\acmConference[iiWAS '17]{The 19th International Conference on Information Integration and Web-based Applications \& Services}{December 4--6, 2017}{Salzburg, Austria}
\acmPrice{15.00}
\acmDOI{10.1145/3151759.3151765}
\acmISBN{978-1-4503-5299-4/17/12}

\begin{document}

\title{Combining Aspects of Genetic Algorithms with Weighted Recommender Hybridization}

\author{Juergen Mueller}
\affiliation{
  \institution{University of Kassel}
  \department{Research Center for Information System Design (ITeG)}
  \streetaddress{Pfannkuchstr. 1}
  \city{34121 Kassel}
  \country{Germany}
}
\email{mueller@cs.uni-kassel.de}

\begin{abstract}


Recommender systems are established means to inspire users to watch interesting movies, discover baby names, or read books. The recommendation quality further improves by combining the results of multiple recommendation algorithms using hybridization methods. In this paper, we focus on the task of combining unscored recommendations into a single ensemble.
Our proposed method is inspired by genetic algorithms. It repeatedly selects items from the recommendations to create a population of items that will be used for the final ensemble. We compare our method with a weighted voting method and test the performance of both in a movie- and name-recommendation scenario.
We were able to outperform the weighted method on both datasets by \HybridMovieLensImprovementVsBaseline and \HybridChallengeImprovementVsBaseline and decreased the overall execution time by up to \HybridMovieLensRuntimeBestShare.
Our results do not only propose a new kind of hybridization method, but introduce the field of recommender hybridization to further work with genetic algorithms.

\end{abstract}

%
%


\maketitle


\section{Introduction}

Recommender systems shape the live on the web on a daily basis. No successful web service runs without them. The hybridization of recommender systems has proven to be effective in addressing the cold start problem or the portfolio effect on all types of recommendation tasks ranging from product recommendations over movie recommendations to name recommendations~\citep{burke2002hybrid,burke2007adaptive,mitzlaff2014summary,harper2015movielens,linden2003amazon,jaeschke2012recommender}.

There is a wide selection of methods to choose from; all with different requirements and purposes. Combining ranked lists of items into a single list is one of those purposes that we want to address in this paper. The task can, broadly speaking, be grouped into two groups: The combination of scored and the combination of unscored rankings. In this paper, we aim at finding a method that is able to efficiently combine two or more unscored source recommendations into a single ranked list of items. Those source recommendations can be without scores or with scores that are deemed not comparable---as it is usually the case.

The most common way to combine multiple recommendations into a single one is the use of a weighted hybridization method. Mostly, it is used with a scored list of recommendations and---in the simplest case---resembles a linear combination of all participating recommendations. \citet{pazzani1999framework} proposed a weighted hybrid that work with unscored recommendations. It treats each recommended item as a vote and the final ensemble is ranked by decreasing vote count. This method works well if there are only a handful of items to recommend from many source recommenders. However, its applicability is limited on long recommendations, because it results in many ties---especially when there are only a handful of combined recommendations. Therefore, we wanted to create a new hybrid method that is able to work even on few large lists of unscored recommended items.

Our proposed novel semi-genetic hybridization method is inspired by genetic algorithms in order to create a new scored list of recommended items from multiple lists of unscored items. It treats the source rankings as chromosome pool where each recommended item is regarded as a chromosome. It then selects a single population from this chromosome pool and ranks those items by their decreasing frequency. The calculated frequency can then be used as scoring for the final ensemble.

The contribution of this paper is as follows: (1) Presentation of a novel semi-genetic hybridization method that is able to efficiently combine two or more long unscored source recommendations into a single ensemble. (2) A comparative experimental study of the quality of the semi-genetic and the weighted hybridization method on two different datasets with four different source recommendations. Experiments on both movie and name data show that this approach is able to outperform a weighted method. We were able to improve the Mean Average Precision~(MAP) score by about \HybridMovieLensImprovementVsBaseline on movie and \HybridChallengeImprovementVsBaseline on name data compared to the weighted method.

The remainder of this paper is structured as follows: The next section gives an overview about the state of recommender hybridization and of genetic algorithms in general. We then describe the proposed semi-genetic hybridization and our baseline---the weighted hybridization using votes---in \secref{method}. The following \secref{data} continues by presenting the used datasets. The core of this paper is found in \secref{experiments}, where the reader finds the results from our experimental study along its discussion. The final section contains a summary of our findings as well as some thoughts for future work.


\section{Related Work}
\label{sec:related-work}


A hybridization method combines the prediction of two or more recommenders to a single recommendation---also called an ensemble. \citet{burke2002hybrid,burke2007adaptive} provides a good survey on hybridization methods. He categorizes the methods into the following seven groups:
\begin{enumerate}

	\item Weighted methods compute the score of a recommended item by adding up the scores from each source recommender. A score of 0 is assumed if an item was not recommended by any source recommender. The simplest approach is to use the linear combination where all recommenders have the same weight~\citep{claypool1999combining}.

	\item Switching methods select one of the given source recommendations to be used based on a given switching criterion. A possible criterion would be to use the secondary recommender if the primary recommender does not provide any items~\citep{tran2000hybrid}.

	\item Mixed methods present the recommendations of multiple recommenders together~\citep{smyth2000personalised}. It is a presentation strategy rather than a combination of multiple recommenders in the sense of obtaining a single list of recommended items.

	\item Feature combination methods are a way to add collaborative filtering data into a content-based recommender by adding the collaborative information as additional features to the data~\citep{basu1998recommendation}.

	\item Cascading methods use one recommendation to rearrange the recommendation of another recommender. A common case would be to order ties where multiple items have the same score~\citep{burke2002hybrid}.

	\item Feature augmentation methods use the output of one recommender to add more information to the data and to be used as additional input for a second recommender~\citep{mooney1999content}.

	\item Meta-level methods are similar to feature augmentation with the difference that the entire model of the first recommender becomes the input for the second one~\citep{balabanovic1998exploring}.

\end{enumerate}
From those hybridization types, the switching, the mixed, the cascading method are not applicable in our scenario, because they are not used to combine multiple recommendations into a single one. Feature combination, feature augmentation, and meta-level methods are extensions of the recommendation algorithms rather than combination strategies for multiple recommendations. Therefore, only the weighted method is applicable in our case.

Recent studies by \citet{paraschakis2016recommender} show that the adoption of sophisticated recommendation algorithms in the industry is rather slow. The authors see the high complexity of most methods as a possible reason. Most hybridization methods on the other hand are rather simple to implement. Therefore, advancing this sub-area of recommender systems could have a higher impact in practice than inventing better and better recommendation algorithms.


\citet{dinoia2014analysis} tested a novel approach on the MovieLens data that brings more diversity to the recommended movies. Their hypothesis is that users who selected diverse items in the past are more likely to want to receive diverse recommendations in the future. Diversity is also increased by applying hybridization methods~\citep{burke2002hybrid}, which will be addressed by this work.

\citet{oconnor2001polylens} proposed the first movie recommender for groups. A key challenge for group recommendations is the creation of one recommendation that suits multiple users. Our semi-genetic hybridization method could be used to combine the single user recommendations of all group members and use the resulting ensemble as group ranking.


Our experiments will also cover the relatively new field of recommending given names. It was firstly proposed by \citet{mitzlaff2013recommending} who compared the performance of a user-based collaborative filtering, item-based collaborative filtering, a weighted matrix factorization method, PageRank, and their newly proposed NameRank using a wide range of evaluation settings. NameRank is an adoption of the PageRank variant proposed by \citet{hotho2006information} who subtracted the global PageRank score from the preferential score. Name recommendation are a relevant task of everyday live, but have shown to be non-trivial to solve.

The name recommendation task was further tested by \citet{mitzlaff2014summary} who conducted a discovery challenge on the task of recommending given names. NameRank was used alongside most popular as baseline during the offline challenge. The challenge confirmed the high difficulty of the task---meaning it is hard to obtain high evaluation scores. Most participants incorporated hybridization as part of their approaches with great success \citep{glauber2014mixed,bayer2014factor,coma-puig2014collaborative,schaefer2014nameling,mitzlaff2014summary}.


Genetic algorithms are population-based search and optimization methods that are inspired by natural evolution~\citep{holland1992adaptation}. They are already widely used in the realm of recommender systems and obtain good results, as shown in the following.

\citet{bobadilla2011improving} proposed a new similarity measure for collaborative filtering that consists of an array of similarity metrics. The authors use genetic algorithms to find ideal weights between those metrics.
\citet{hwang2010using} and \citet{salehi2013hybrid} have done a similar work where they use genetic algorithms to weight the feature vectors that is used to compute the similarity between users.

\citet{kim2005using} presented a genetic algorithm-based k-Means algorithm that separates the user-base before a recommender is applied. The genetic algorithm was used to find the initial seeds for the clustering. Doing so enables a more targeted use of recommenders.

\citet{georgiou2010improving} developed a whole clustering algorithm that is based on genetic algorithms. In their approach, each chromosome is a randomly created clustering---it defines the users that belong to the cluster and those that do not--and their algorithm tries to increase the number of users in the cluster while increasing the number of shared items.


\section{Methods}
\label{sec:method}
This section presents the problem to be solved as well as the method we use to address it. This constitutes of our novel semi-genetic hybridization method as well as the weighted hybridization voting method by \citet{pazzani1999framework}, which will serve as baseline during our experiments.

\subsection{Problem Definition}
During our experiments, we deal with a standard binary item recommendation task. Users of a web service have expressed interest in certain items by requesting them. These interactions with the system are interpreted as (binary) positive feedback to these items.

We use two publicly available datasets: The MovieLens data about movie ratings and the Nameling discovery challenge data about name requests \seesec{data}. For each user, the last two items from the usage history are left-out. The first will be used for training and the second for evaluation. The task during the experiment is then to produce for each user a list of $k$ item recommendations, ordered by their relevance to the user at hand.

We select four source recommender that should be combined using the hybridization method at hand. The obtained ensemble should improve the performance of the best single recommender while requiring as little additional runtime as possible.

\subsection{The Semi-Genetic Hybridization Method}
The basic idea of our semi-genetic hybridization method is to use aspects of genetic algorithms to create the ensemble. Our goal is to use those aspect to create a non-linear combination of all source recommendations while honoring the inherent item-relevancy given by the items ranking.

Genetic algorithms in general are stochastic search and optimization algorithms that mimic the natural selection to find a good solution. Given a so-called chromosome pool of items to draw from, they typically follow the following steps to obtain this result:
\begin{enumerate}

	\item Initialize Population: A first population of chromosomes is created---usually at random---from the chromosome pool.

	\item Evaluate Fitness: During each iteration, the fitness of each chromosome is evaluated using a given fitness function.

	\item Selection: A subset of the population is selected at random based on their fitness for mating. It is possible that any given item is selected multiple times.

	\item Cross-over: Each pair of chromosomes is mated using a cross-over technique that exchanges parts of the chromosome's information.

	\item Mutation: Additionally, a mutation can be applied with a very low chance that alters part of the chromosomes or replaces them altogether by new ones.

	\item Solution Set: The genetic algorithm stops if a certain criterion is reached. The simplest one would be a pre-defined number of iterations. The last population that remains after the termination---called the solution set---is used to solve the task at hand.

\end{enumerate}

Typically, genetic algorithms run for multiple iterations until a certain fitness quality is reached. This behavior can result in a high demand for memory and runtime. However, a hybridization method should add as little runtime to the recommendation process as possible, which is in stark contrast to the way genetic algorithms usually operate. Therefore, a balance between a high performance and a low runtime.

Before genetic algorithms can be applied, one has to define the following features: A suitable encoding for the chromosomes and a fitness function. The fitness function is a central feature, because it decides which chromosomes survive a given iteration. It should reflect the chance of the given chromosome to solve the problem at hand. However, there is a reason that genetic algorithms have not already been applied to recommender systems: There is no such function. A list of recommended items contains some that hit the users taste and others that do not, but none in-between. So, how to assign a useful fitness score?

The situation is a little different for the hybridization task. There is an implicit score attached to each item in the source recommendations: It's position in the ranking. It can safely be assumed that any recommendation algorithm puts more promising items first. Therefore, we decided to use each recommended item as a chromosome and evaluate its fitness by using its reciprocal rank. The fitness score is computed as follows:

\begin{equation}
	\text{ReciprocalRank}(i) = \frac{1}{\text{Rank}(i)},
	\label{eq:reciprocal-rank}
\end{equation}
where $i$ is an item from a recommended list of items $R$ and Rank($i$) is the rank position of $i$ in $R$.

However, this implicit relevancy score is only present in the first iteration where we can use the output of the recommendation algorithms directly. We lose this information afterwards. Furthermore, there is no meaningful way to apply any cross-over to the chromosomes when they are the smallest entity of the task (in our scenario: a recommended item). Therefore, we refrain from applying any cross-over or mutation. We further refrained from doing more than one iterations, which makes our approach a single pass process.

Avoiding cross-over, mutations, and iterations has a beneficial side-effect for our hybridization task: It keeps the additional runtime low. The lack of those parts of genetic algorithms lead to our decision to call our hybridization methods semi-genetic, because it uses only aspects of them.

\figref{hybrid-ga-vs-hybrid} shows the proposed mapping from genetic algorithms to our semi-genetic hybridization. It basically, contains only the first half of the genetic algorithm steps in favor of a low runtime.

\begin{enumerate}

	\item Initialize Population (input $\mathcal{R}$ in \srcref{genetic-hybrid}): We treat the source recommendations from the recommender systems that should be combined as population initialization. They basically select a set of items from all possible chromosomes.

	\item Evaluate Fitness (\srcref{genetic-hybrid}, line \ref{code:genetic-hybrid-evaluate-fitness-start} - \ref{code:genetic-hybrid-evaluate-fitness-stop}): The fitness of the recommended items is evaluated by assigning the reciprocal rank to them.

	\item Selection (\srcref{genetic-hybrid}, line \ref{code:genetic-hybrid-selection}): We then use the selection step to create a huge population of items by randomly drawing $n$ items with replacement from the initial population based on their fitness score (i.e., reciprocal rank).

	\item Cross-over: None.

	\item Mutation: None.

	\item Solution Set (\srcref{genetic-hybrid}, line \ref{code:genetic-hybrid-solution-set-start} - \ref{code:genetic-hybrid-solution-set-stop}): The method terminates after the first iteration and the resulting solution set is used for the combined ranking by counting the frequency of each item and ordering them by decreasing frequency.

\end{enumerate}

\fig{}{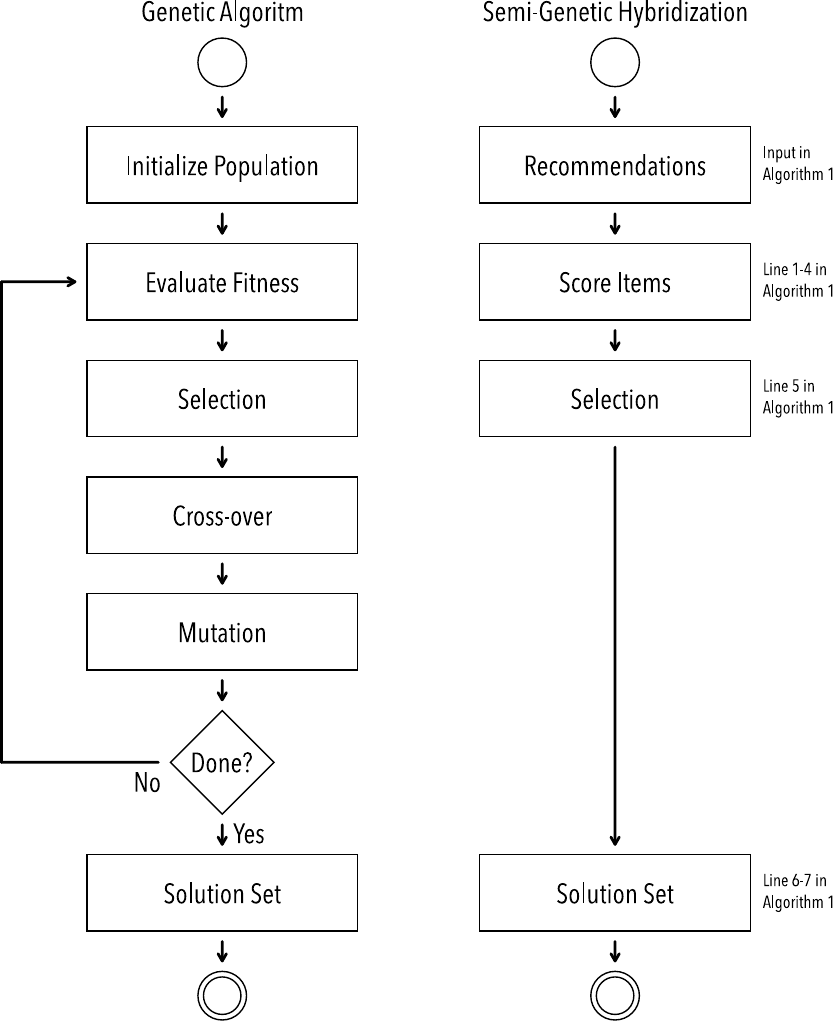}{Comparison between the workflow of a regular genetic algorithm and our proposed semi-genetic hybridization method.}

\srcref{genetic-hybrid} shows the methods workflow as a pseudo-code. The hybrid gets the collection of recommendations $R$ to be hybridized and a numeric value $n$ given the number of random samples to draw from the recommendations---which is the first and final iteration. One item can occur in multiple source recommendations with differing fitness values. This results in a higher survival rate for the given item, because it increases its chance to be drawn.

\begin{algorithm}
	\KwData{A collection of recommendations $\mathcal{R}$ consisting of two or more recommendations $R \in \mathcal{R}$ containing $|R|$ ranked items $i \in R$.}
	\KwResult{A list of ranked items that contains items of $\mathcal{R}$.}

	\tcp{Evaluate fitness / score items}
	\ForEach{recommendation $R \in \mathcal{R}$}{\label{code:genetic-hybrid-evaluate-fitness-start}
		\ForEach{item $i \in R$} {
			Assign the reciprocal rank to $i$ as fitness score;
		}
	}

	Create a multi set $P$ from all $r \in R$;\label{code:genetic-hybrid-evaluate-fitness-stop}

	\tcp{Selection}
	Create a multi set $P'$ by drawing randomly $n$ items with replacement from $P$ using the fitness score as probability;\label{code:genetic-hybrid-selection}

	\tcp{Solution set}
	Count the frequency of each item $i \in P'$;\label{code:genetic-hybrid-solution-set-start}

	Order the items by decreasing frequency;\label{code:genetic-hybrid-solution-set-stop}

	\caption{Semi-genetic hybridization method}
	\label{code:genetic-hybrid}
\end{algorithm}

Our approach has the advantage of being applicable on any list of recommended items---with or without scores. The fitness function can be replaced by the item score if the source recommender has scores that are deemed comparable.

\subsection{The Weighted Method}

We select the weighted hybridization method that was proposed by \citet{pazzani1999framework} as our baseline method during our experiments, because it is a well-established method.

The weighted hybridization mechanism for unscored items is basically a voting system where each recommender can put $n$ votes---at most one vote per items---where $n$ is the number of recommended items. The items are then ordered by the decreasing number of votes. Ties are solved using the order in the source recommendations.

This method has the benefit that it can work with recommended items that have no score attached or in cases where the scores of different recommenders mean vastly different things. However, this method results in many ties in cases where there is a huge number of votes to give per recommender with only a small number of participating recommenders. The weighted voting method is described in \srcref{baseline} as pseudo-code.

\begin{algorithm}
	\KwData{A collection of recommendations $\mathcal{R}$ consisting of two or more recommendations $R \in \mathcal{R}$ containing $|R|$ ranked items $i \in R$.}
	\KwResult{A list of ranked items that contains items of $\mathcal{R}$.}

	\ForEach{recommendation $R \in \mathcal{R}$}{
		\ForEach{item $i \in R$} {
			Assign a voting weight of $1$ to $i$.
		}
	}

	Create a multi set $P$ from all $r \in R$;

	Count the number of votes of all $i \in P$ (i.e., the frequency of each item);

	Order the items by decreasing frequency;

	\caption{Weighted hybridization method using votes}
	\label{code:baseline}
\end{algorithm}

The core difference to our semi-genetic hybridization method is to be found in two crucial steps: It has no selection step and the solution set is created differently. Our semi-genetic method counts the number of items (i.e., chromosomes) not their fitness score; the weighted method counts the vote scores of the items---which corresponds to counting the items in our case, given that each item has a voting score of 1.

\section{Data}
\label{sec:data}
This section presents the data that will be used during our experiments. We use two datasets, one containing movie ratings from MovieLens, one containing name searches from the Nameling name search engine.

\subsection{MovieLens 100K Data}
MovieLens is a well-known service for personal movie recommendations by GroupLens\furl{https://movielens.org}---a research lab at the University of Minnesota. It offers rich information and enables its users to find, tag, and rate movies. GroupLens has published multiple datasets from their web system for scientific use.

We will use the well-known MovieLens 100K dataset\fsource{File ``ml-100k.zip'' from \url{https://grouplens.org/datasets/movielens/100k}}~\citep{harper2015movielens} during our experiments. This dataset contains 99392~ratings for 1664~movies from 943~users with 20 movie ratings per user (users with less movie rating where removed from the dataset by GroupLens).

\subsection{Nameling Discovery Challenge Data}

\begin{figure}[t]
  \centering
  \includegraphics[width=0.7\columnwidth]{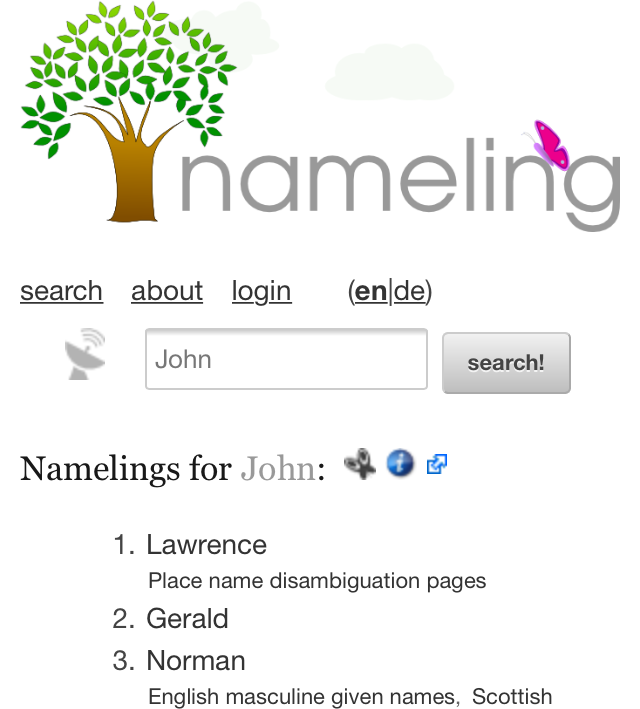}
  \caption{Screenshot of the Nameling web application showing the first three search results for a the name ``John''.}
  \label{fig:hybrid-nameling-screenshot}
\end{figure}

Nameling \citep{mitzlaff2012namelings} is designed as a search engine and recommendation system for given names. The basic principle is simple: The user enters a given name and gets a browse-able list of relevant, related names. As an example, \figref{hybrid-nameling-screenshot} shows the similar names for the classical masculine American given name ``John''.\furl{http://nameling.net/name/John} The list of similar names in this example (``Lawrence'', ``Gerald'', and ``Norman'') exclusively contains classical English masculine given names as well. Categories for the respective given name are displayed whenever such a Wikipedia article exists (e.g., ``English masculine given names'', ``Scottish masculine given names'', and ``Surnames'' for the given name ``Norman''). Via hyperlinks, the user can browse for similar names of each listed name or get a list of all names linked to a certain category in Wikipedia.

We use the publically available offline challenge data\fsource{File ``20DC13\_Offline\_Challenge.tar.bz2'' from \url{https://kde.cs.uni-kassel.de/nameling/dumps}} from the Nameling discovery challenge as well as the list of submitted name recommendations from the top~6 teams.\fsource{File ``20DC13\_Top6\_Submissions.tar.bz2'' from \url{https://kde.cs.uni-kassel.de/nameling/dumps}} We use the list of recommended names that where originally submitted by the participating teams as source recommendations during our hybridization experiments. The dataset contains 515848 activities from 60922 users in the train data and two left-out names for \HybridChallengeNumUsers test users each in the test data. Each submitted name recommendations contains 1000 names for each of the \HybridChallengeNumUsers test users.


\section{Testing the Methods}
\label{sec:experiments}
This section presents the experimental design, data pre-processing, evaluation metric, describes the obtained results, and the discussion of those results.

\subsection{Design and Procedure}
Our experiments focus on the task of combining recommendations that were made by several recommenders beforehand. Our novel semi-genetic hybridization method will be compared against the weighted voting method by \citet{pazzani1999framework} as well as the best single result of a source recommender.

The hybridization methods will further be evaluated based on their runtime, because most recommendation scenarios require fast responses and adding hybridization increases the overall runtime anyhow. Therefore, we need to find a suitable upper boundary for the usable runtime. We will further address this topic in \secref{experiments-parameter-optimization}.

All experiments were conducted using R\furl{https://r-project.org} in version~3.3.3 and the MovieLens recommendations were created using the recommenderlab r-package\furl{https://cran.r-project.org/web/packages/recommenderlab} in version 0.2-2. We ensured that we had no concurrent processes running on the test cluster to ensure comparable runtime measurements between the semi-genetic and weighted method.

\subsection{Evaluation}
Given the list of test users and a set of source recommendations, each method produces a list of $k$ recommended items per user. These lists are then used to evaluate the quality of the method by comparing for each test user the $k$ items to the left-out items from the test data. It is assumed that good methods will rank the left-out items high, since they represent the actual measurable interests of the user at hand.

The chosen assessment metric to compare the lists of recommendations is Mean Average Precision~(MAP). MAP can usually handle arbitrarily long lists of recommendations; however, we restricted it to MAP@1000, because this is the number of items that were recommended in the discovery challenge data, meaning that only the first 1000 positions of a list are considered. More formally, the score assigned to a recommended list is
\begin{equation}
   \text{MAP@k} = \frac{1}{|U|}\sum_{u = 1}^{|U|}\frac{\left(\frac{1}{\text{Rank}(u_{i_1})} + \frac{2}{\text{Rank}(u_{i_2})}\right)}{2},
\end{equation}
where $U$ is the set of all test users, Rank($u_{i_1}$) and Rank($u_{i_2}$) are the ranks of two left-out items for user $u \in U$ from the secret evaluation dataset. Rank($u_i$) returns 0 if an item does not occur in the top $k$ list positions of recommended items.

\subsection{Data Pre-Processing}
We conducted the following pre-processing, in order to create a comparable setting between the MovieLens and Nameling data:

We need to have three datasets: Train data to create the source recommendations, tuning data to optimize the model parameter, and testing data to evaluate the final performance.

In order to obtain this setting in the MovieLens data, we use the last movie rating as testing data, the rating next to the last as tuning data, and the remaining ratings as training data.

The Nameling data is already separated into training and test data. The train data was created be striping the data starting with the next to last name search requests for \HybridChallengeNumUsers users. The last two name search request are provided as test data. We further divide this original test data into tuning and testing by using the first name as training and the second as testing name.

Next, we need a set of source recommendation that can be used for hybridization. For the Nameling data, there is already a set of the best recommendations of the Nameling discovery challenge as download available. We will use them as non-trivial source recommenders. For the MovieLens data, we use the common User-Based Collaborative Filtering~(UBCF), Singular Value Decomposition~(SVD), Item-Based Collaborative Filtering~(IBCF), and a Most Popular recommender to form the source recommendations using the train data.

\tabref{hybrid-results-ranking-movielens} and \ref{tab:hybrid-results-ranking-challenge} show the performance of all source recommenders by decreasing MAP@1000 score. During our experiments, we will combine between two and four of those source recommenders using their decreasing MAP@1000 score order. We will label them Top~2, Top~3, and Top~4 for a combination of the first two, three, and four best (e.g., the Top~3 ensemble on the Nameling data is a hybridization of uefs.br, ibayer, and all\_your\_base).


\begin{table}[tb]
	\centering
	\caption{The results for the four methods on the MovieLens data, showing the method names together with their achieved MAP@1000 score.}
	\label{tab:hybrid-results-ranking-movielens}
	\begin{tabular}{rlr}
		\toprule
		Pos. & Method                                               &  Score \\
		\midrule
		   1 & User-based Collaborative Filtering (UBCF)            & 0.0309 \\
		   2 & Most Popular                                         & 0.0295 \\
		   3 & Singular Value Decomposition (SVD)                   & 0.0198 \\
		   4 & Item-based Collaborative Filtering (IBCF)            & 0.0126 \\
		\bottomrule
	\end{tabular}
\end{table}



\begin{table}[tb]
	\centering
	\caption{The results for the four best teams of the Nameling discovery challenge, showing the team names together with their achieved MAP@1000 score.}
	\label{tab:hybrid-results-ranking-challenge}
	\begin{tabular}{rlr}
		\toprule
		Pos. & Team Name                                                               &  Score \\
		\midrule
		   1 & \citet{glauber2014mixed} (uefs.br)                                      & 0.0486 \\
		   2 & \citet{bayer2014factor} (ibayer)                                        & 0.0419 \\
		   3 & \citet{letham2014similarity} (all\_your\_base)                          & 0.0418 \\
		   4 & Aurélio Domingues et al. \citep{aureliodomingues2014improving} (Labic)  & 0.0374 \\
		\bottomrule
	\end{tabular}
\end{table}

\subsection{Parameter Optimization}
\label{sec:experiments-parameter-optimization}
First, we need to identify a suitable value for the parameter $n$ for the size of the population for our semi-genetic hybridization method before we can compare our semi-genetic hybridization with the weighted method. We conducted a grid-search for $n$ on the tuning data where we tested values ranging from 1000 to 40000 in steps of 1000. We further repeated every run \HybridMovieLensRepeats times in order to account for the randomness in the semi-genetic hybridization method. We also run each weighted hybridization experiments \HybridMovieLensRepeats times to obtain reference runtime measurements of the hybridization methods themselves.

\begin{figure*}[!t]
   \centering

   \subfig{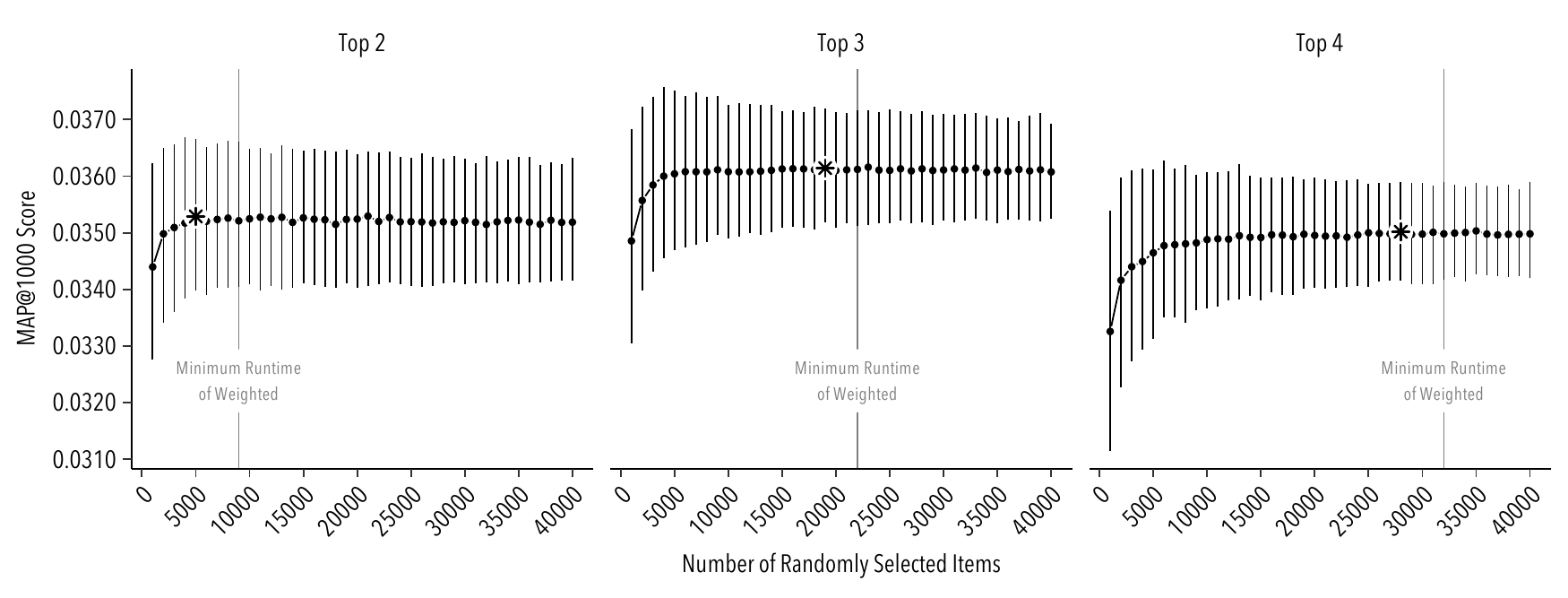}{MovieLens data}

   \subfig{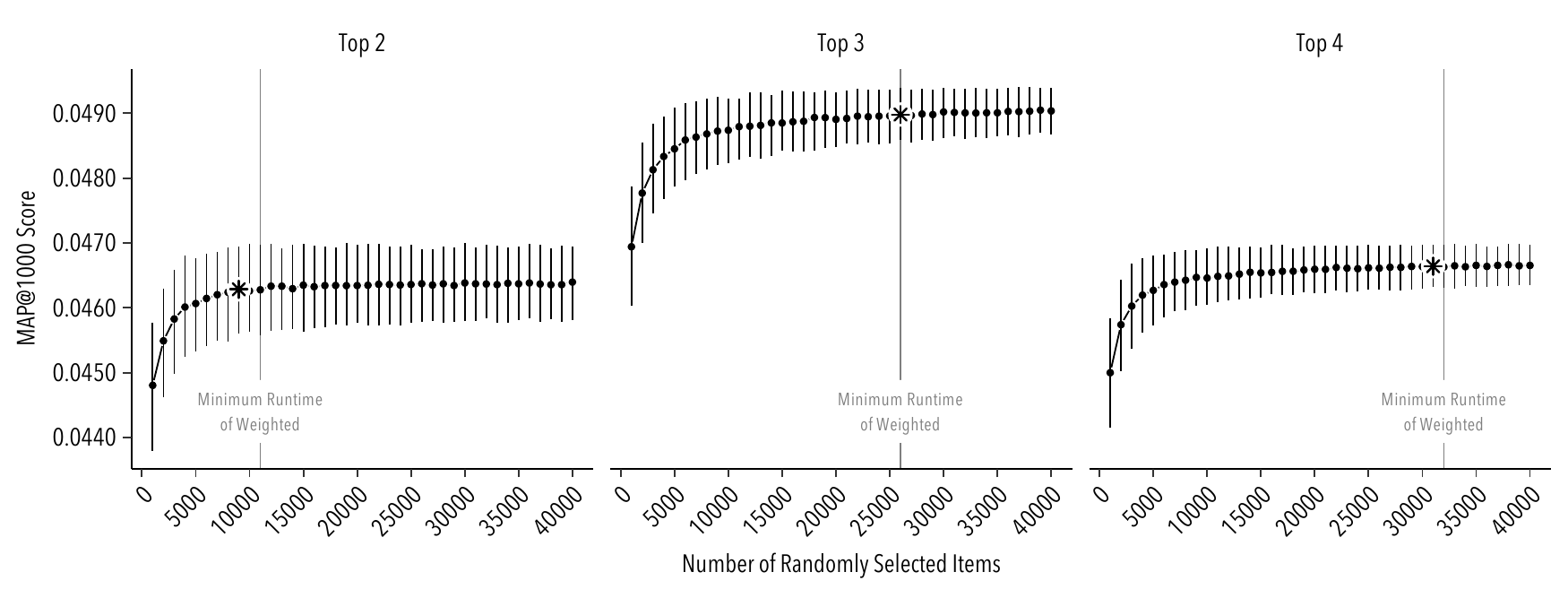}{Nameling data}

   \caption{Median MAP@1000 scores for the grid-search for the number of randomly selected items with three different hybridizations per dataset and \HybridMovieLensRepeats repeats on the tuning data. Those runs that obtained the highest scores while running faster than the fastest baseline are marked with a \ding{83}-label. Note that the y-axis does not start a 0 to better illustrate the differences. Error bars show 95~\% confidence intervals.}
   \label{fig:hybrid-results-grid-search}
\end{figure*}

\figref{hybrid-results-grid-search-movielens} show the tuning results for the MovieLens data and \figref{hybrid-results-grid-search-challenge} shows the results for the Nameling data.

The plots in \figref{hybrid-results-grid-search} show the median scores across all \HybridMovieLensRepeats runs as line with error bars showing the 95~\% confidence interval. The vertical line in the plots mark the slowest experiment that is still faster or as fast as the fastest baseline run.

Most combinations follow the same course with a steep increase in the MAP@1000 score for increasing values of $n$ with a plateauing effect at a size of about 10000. Note that the actual MAP@1000 score does not matter during this experiment; only its change compared to other parameters.

The runtime increases with an increasing value of the number of randomly selected items. Choosing simply the best MAP@1000 score would ignore the runtime as a factor. A hybridization technique always increases the overall runtime of a recommendation, but should require as little runtime as possible. So, what is a reasonable upper threshold considering the runtime? We decided to use the minimum runtime of the weighted hybrid method as an upper bound for the semi-genetic method. The weighted method is accepted in practice and, consequently, deemed fast enough. Therefore, the semi-genetic method can be considered fast enough, as long as it runs as fast as or faster than the fastest weighed method. Therefore, we selected the best median MAP@1000 score for every hybridization running faster than the fastest weighted method. This leads to a good balance of MAP@1000 score and runtime while staying at worst as fast as the baseline. Doing so, we selected $n = \HybridMovieLensParamTopTwo$ for the Top~2, and $n = \HybridMovieLensParamTopThree$ for the Top~3 hybrid in the MovieLens experiments and $n = \HybridChallengeParamTopTwo$ for the Top~2, and $n = \HybridChallengeParamTopThree$ for the Top~3 hybrid in the Nameling experiments. The selected parameters for $n$ are marked with a \ding{83}-label in \figref{hybrid-results-grid-search}.

\subsection{Hybridization Evaluation}
We conducted our evaluation on the testing data with the abovementioned parameters that we found during the grid-search for each Top~$n$ hybridization. The task during our experiment is to predict the left-out items of each user in the testing data.

\figref{hybrid-results-evaluation-movielens} shows the obtained median results for each ensemble on the MovieLens data as well as the best score that was obtained by the best single recommender (i.e., UBCF with a MAP@1000 score of \HybridMovieLensScoreWinner).

All hybridization result follows the same trend across both methods: The score decreases with an increasing number of combined recommenders. In other word, combining more and more source recommenders does not benefit the overall performance, which is to be expected, because every additional source recommender has a lower MAP@1000 score than the previous ones. Our proposed semi-genetic hybridization outperforms the weighted hybrid in every experiment by about \HybridMovieLensImprovementVsBaseline on average. Further, does the Top~2 outperform the singe recommender with \HybridMovieLensScoreBest by about \HybridMovieLensImprovementVsWinner in the Top~2 experiment. The Top~3 ensemble obtains a slightly better result than the best single recommender and the Top~3 ensemble is slightly worse. The weighted method is only able to obtain slightly better results than the best single recommender with the Top~2 ensemble and is clearly inferior than the best single recommender in the Top~3 and Top~4 experiment. This dramatic decrease in MAP@1000 scores can be explained by the sharp decrease in the MAP@1000 scores of the added source recommenders as given in \tabref{hybrid-results-ranking-movielens}.

\begin{figure}[t]
   \centering

   \subfig{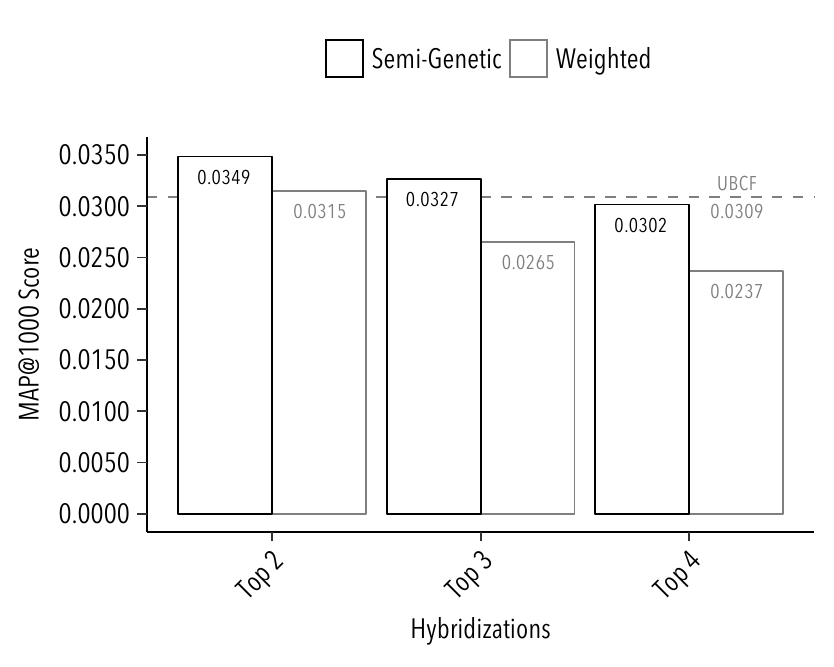}{MovieLens data}

   \subfig{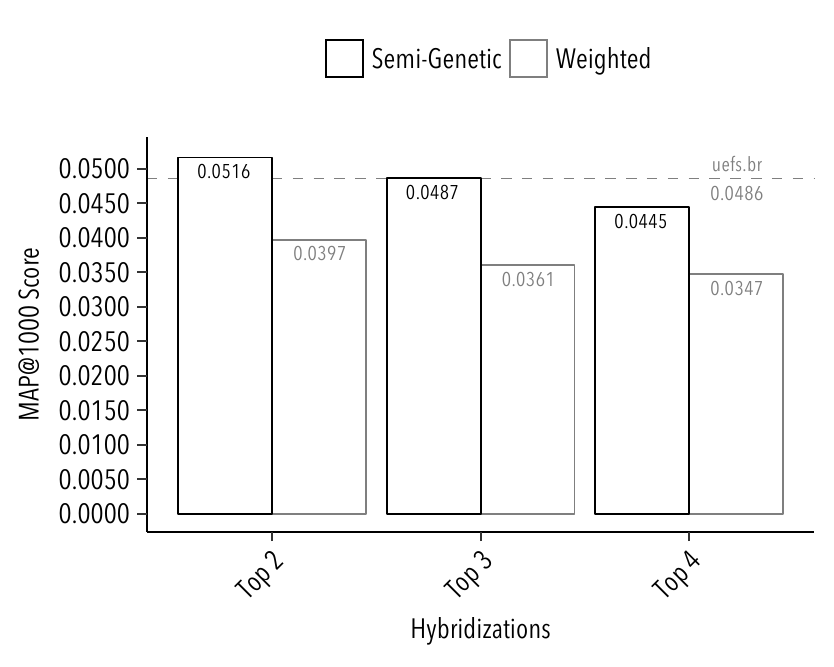}{Nameling data}

   \caption{Median MAP@1000 scores on the test data with \HybridMovieLensRepeats runs across all hybridizations. The horizontal line marks the best obtained result from the sole recommendations.}
   \label{fig:hybrid-results-evaluation}
\end{figure}

\figref{hybrid-results-evaluation-challenge} shows the obtained median results for each ensemble on the Nameling data as well as the best score that was obtained by the best team (i.e., team uefs.by with a MAP@1000 score of \HybridChallengeScoreWinner).

The results follow the same trend as on the MovieLens data: Decreasing scores with larger ensembles. Again, our proposed semi-genetic hybridization outperforms the weighted hybrid in every experiment---in this case by about \HybridChallengeImprovementVsBaseline. The Top~2 outperform the best team with \HybridChallengeScoreBest by about \HybridChallengeImprovementVsWinner in the Top~2 experiment. The Top~3 ensemble obtains about the same results as the best team and the Top~4 ensemble is slightly worse. The downward trend with the increasing number of combined teams can be explained analogue to the MovieLens experiment with the decreasing MAP@1000 score of the combined recommendations. The decline is less pronounced, because the MAP@1000 scores of the source recommendations are closer to each other \seetab{hybrid-results-ranking-challenge}. The weighted method performs very different than on the MovieLens data---it is never able to obtain better results than the best single recommendation.

\begin{figure}[t]
   \centering

   \subfig{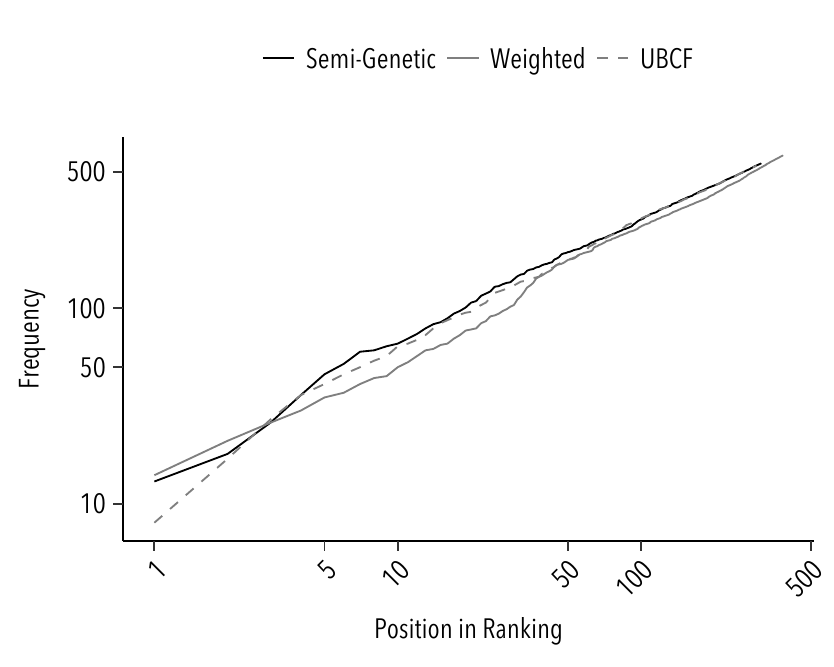}{MovieLens data (\HybridMovieLensLeftOutItems left-out movies)}

   \subfig{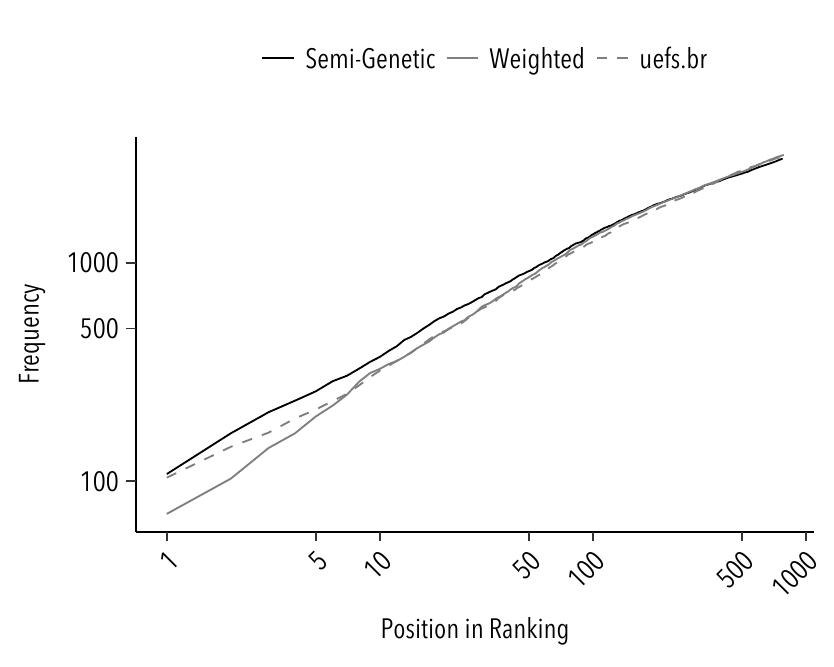}{Nameling data (\HybridChallengeLeftOutItems left-out names)}

   \caption{Cumulative distribution of the ranking positions for both hybrid methods as well as the respective winners (i.e., User-based Collaborative Filtering (UBCF) and uefs.br). Both axes are logarithmically scaled.}
   \label{fig:hybrid-rank-positions}
\end{figure}

To compare the recommenders' performances in greater detail, \figref{hybrid-rank-positions} shows the cumulative distribution of the different ranking positions ($1, \ldots, 1000$). For each recommender system and every ranking position $k$ is the number of hold-out items displayed---from the test users with one hold-out item each---that had a rank smaller than or equal to $k$ on the list of recommended items.

\figref{hybrid-results-rank-positions-movielens} shows the ranking position on the MovieLens data for the Top~2 hybridization for the semi-genetic method, weighted method, and the best single recommender (i.e., UBCF). We can see a clear difference for the rank position in the first 50 positions. Both hybridization methods are able to increase the number of movies on the top three positions compared to UBCF. The weighted method is even able to recommend a few more movies at those positions than our semi-genetic method. However, it falls behind shortly after when both UBCF as well as our semi-genetic method place far more movies correctly between position five and fifty with the semi-genetic method being always better than UBCF. Both, UBCF and the semi-genetic method are about equal starting from position fifty with the weighted method always staying behind. However, the weighted method is able to recommend more movie correctly within the 1000 items, but at very low positions.

\figref{hybrid-results-rank-positions-challenge} shows the ranking position on the Nameling data for the Top~2 hybridization for semi-genetic method, weighted method, and the best single recommender (i.e., uefs.br). We see that the challenge winner has predicted about as much names correctly onto position one as our semi-genetic method. However, the semi-genetic method is able to place more names in high positions for about the first 100 positions. The weighted method on the other hand, obtains far lower results than team uefs.br or the semi-genetic method, which is to be expected given the low MAP@1000 scores.

To wrap up our analysis: Our goal was to find a better hybridization method to be applied on long non-scored rankings. We used state of the art recommendation methods for the MovieLens data and the predictions that were made by the top participants of the Nameling discovery challenge for the Nameling data. We further restricted our semi-genetic method to require no more runtime than the fastest weighed method run.
Doings so, we were able to outperform the weighted baseline for all Top~$n$ hybridizations. The improvement over the baseline is quite pronounced across all experiments. To further illustrate this difference using the Top~2 ensemble: The improvement corresponds to an average increase of a left-out item from about position 32 to 29 for the MovieLens data and to an increase from position 25 to 20.

\subsection{Runtime Analysis}
The second aspect of our hybridization experiments is the impact on the runtime. We selected the number of randomly selected items in a way to ensure that our semi-genetic method adds at worst about the same amount of time to the overall recommendation as the weighted method. Please note that the following experiments focus on the runtime of the hybridization methods themselves and do not take the runtime of the combined source recommendations into account.

\begin{figure}[t]
   \centering

   \subfig{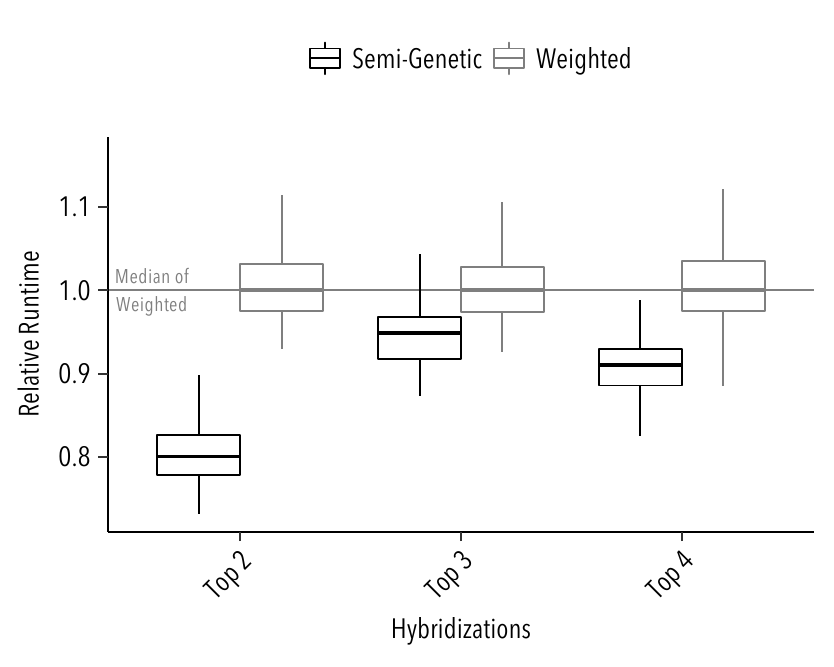}{MovieLens data}

   \subfig{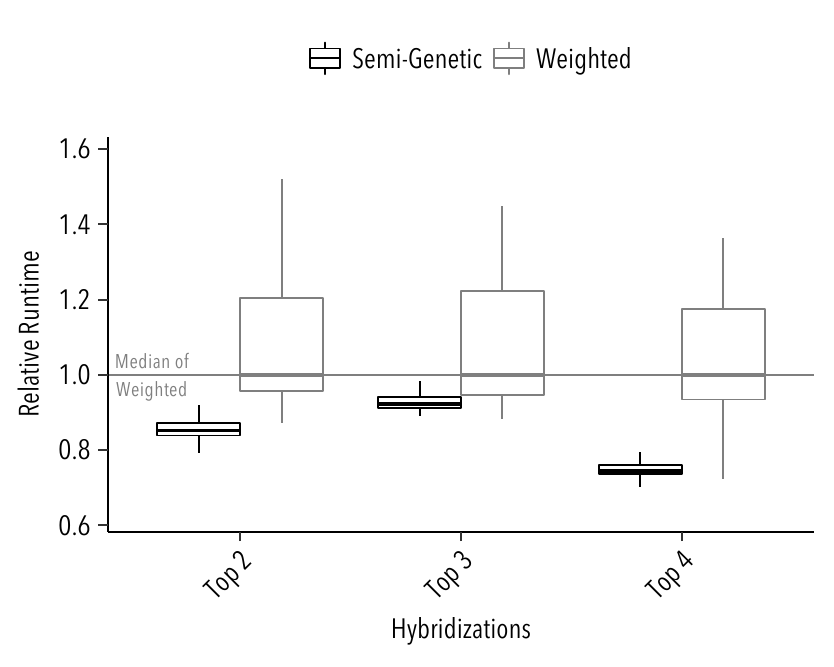}{Nameling data}

   \caption{Boxplot showing the execution time on the testing data with \HybridMovieLensRepeats repeats in each experiment. The runtime is normalized to the median runtime of the weighted method better to illustrate the difference between both methods.}
   \label{fig:hybrid-results-runtime}
\end{figure}

The measured results shown in \figref{hybrid-results-runtime} show that our selection technique indeed resulted in comparable or better runtimes of the semi-genetic over the weighted method in all experiments.
In order to better illustrate the difference between both method, we decided to compare each experiment's runtime with the Median runtime of the weighted method---meaning, the plot shows how much faster/slower the semi-genetic method is compared to the Median weighted method in a given Top~$n$ ensemble.

The Top~2 ensemble run significantly faster than the weighted method by up to \HybridMovieLensRuntimeBestShare. This is due to the fact the selected population size is with \HybridMovieLensParamTopTwo very small. We further see that---except of some runs in the highest quartile for the Top~3 ensemble---all experiments run faster than the lower quartile runtime of the weighted method.

The results on the Nameling data are similar, with the Top~2 ensemble being up to \HybridChallengeRuntimeBestShare faster than the Median weighted method. As on the MovieLens data, all experiments run bellow the lower quartile runtime of the weighted method experiments.

In summary, our results show that our semi-genetic method is not just better in terms of MAP@1000 score, but is also able to run faster than the weighted baseline. This difference is to be found across all experiments, which leads to the conclusion that our semi-generic hybridization method is a suitable replacement for the weighted voting method.


\section{Conclusions}

In this paper, we presented a hybridization method for recommender systems that picks up aspects of genetic algorithms. The combined recommendation is constructed by randomly selecting items with replacement from the source recommendations based on the reciprocal rank. It provides a significant improvement in scenarios where long lists of recommended items offer no scores or scores that are deemed to be not comparable.

Our experiments show a clear improvement of up to \HybridChallengeImprovementVsBaseline (on the Nameling data) over the traditional weighted method, which uses the items in the source recommendations as votes. This is especially difficult when there are only a few recommenders with lots of items that should be combined, because there are not enough vote ranks---the number of source recommenders at most---compared to the number of items.

A possible extension to our approach would be to use different weights for each recommender and each user. We were not able to do this experiment, because one would need to repeat the experiment over time on each user multiple time to learn the preferences of that particular user in an online evaluation. However, our experiments relied on offline data, which do not offer this opportunity.


\section{Acknowledgments}
We thank Stephan Doerfel, Robert Jäschke, Bastian Schäfermeier, Christoph Scholz, and Gerd Stumme for comments that greatly improved the manuscript.

\bibliographystyle{ACM-Reference-Format}
\bibliography{hybridization-paper}
\end{document}